\documentclass[apjl]{emulateapj}

\newcommand{\etal}{et al.}
\newcommand{\sgra}{Sgr A*}
\newcommand{\poincare}{Poincar\'{e}}

\shorttitle{RIAFs and Polarization}
\shortauthors{Ballantyne \etal}

\usepackage{times}

\begin{document}

\title{Constraining Radiatively Inefficient Accretion Flows with Polarization}


\author{D. R. Ballantyne\altaffilmark{1}, Feryal \"{O}zel\altaffilmark{1}, and Dimitrios
  Psaltis\altaffilmark{1,2}}
\altaffiltext{1}{Department of Physics, The University of Arizona, 1118 East 4th
  Street, Tucson, AZ 85721; drb, fozel, dpsaltis@physics.arizona.edu}
\altaffiltext{2}{Also at: Department of Astronomy, The University of Arizona, 933
  N. Cherry Avenue, Tucson, AZ 85721}

\begin{abstract}
The low-luminosity black hole \sgra\ provides a testbed
for models of Radiatively Inefficient Accretion Flows (RIAFs). Recent
sub-millimeter linear polarization measurements of \sgra\ have
provided evidence that the electrons in the accretion flow are
relativistic over a large range of radii. Here, we show that
these high temperatures result in elliptical plasma normal
modes. Thus, polarized millimeter and sub-millimeter radiation emitted
within RIAFs will undergo generalized Faraday rotation, a cyclic
conversion between linear and circular polarization. This effect will
not depolarize the radiation even if the rotation measure is
extremely high. Rather, the beam will take on the linear and circular
polarization properties of the plasma normal modes. As a
result, polarization measurements of \sgra\ in this frequency regime
will constrain the temperature, density and magnetic profiles of RIAF
models.
\end{abstract}


\section{Introduction}
\label{sect:intro}
Radiatively Inefficient Accretion Flows (RIAFs) are models of black
hole accretion at rates much smaller than
the Eddington rate. In this regime, numerous physical effects reduce the ability of the infalling gas to radiate, thereby robbing
accretion of its capacity to convert gravitational energy into
light. For example, in an early RIAF model known as an Advection
Dominated Accretion Flow (ADAF) the gas density is
low enough that the electrons and ions thermally decouple
\citep[e.g.,][]{sle76}, so that the accretion energy is carried
by the ions into the black hole with little escaping as radiation
\citep{ich77,ree82,ny95}. Later work found that these low
density flows resulted in gas that is unstable to large radial circulations
\citep{qg00a,ian00} and only marginally bound to the black hole
\citep{bb99}. Thus, given a tiny perturbation, only a fraction of the gas that is fed to
the outer regions of the accretion flow will actually successfully
fall into the black hole, further reducing the radiative efficiency.

The benchmark against which all RIAF models are tested is the
black hole at the center of the Galaxy, \sgra, which
radiates at $3\times 10^{-9}$ of its Eddington
luminosity, and has a well determined spectral energy distribution from
radio wavelengths to X-ray energies in both quiescent and flaring
states \citep{mf01,bag03,gen03}. Several different RIAF models fit the spectrum of \sgra\ \citep[e.g.,][]{nym95,qn99,opn00,yqn03},
  resulting in different solutions to the desired physical
  parameters such as the run of temperature, magnetic field strength, and
  density with radius. The constraints tightened when
  \citet{ait00} reported the tentative detection of
  linear polarization in the sub-millimeter spectrum of
  \sgra. \citet{agol00} and \citet{qg00} then showed that the
  higher density ADAF solutions would likely depolarize any emission
  in this wavelength range, and speculated that only the hotter, more tenuous
  models are viable candidates. Now, with several successful
  observations, linear polarization measurements are being used to constrain the
  accretion rate and density profile of the flow onto \sgra\ \citep[e.g.,][]{mar06,mar07}.

This paper discusses how the relativistic plasma in RIAF solutions will effect the
polarization properties of any propagating radiation. In particular, we
show that observations of both
linear and circular polarization at multiple frequencies can provide important constraints on many of the key predictions of RIAFs.

\section{Rotation and Conversion of Polarized Light}
\label{sect:plasma}
Radiation propagating through a plasma pierced by a uniform external
magnetic field will travel via two wave modes that have orthogonal
polarizations and different phase velocities. The polarization
properties of the radiation are thus altered as it passes through the
medium. The best known example of this effect is Faraday rotation,
where a linearly polarized wave has its plane of polarization
rotated during passage through a cold (i.e., $\theta \ll 1$, where
$\theta \equiv kT_e/mc^2$ and $T_e$ and $m$ are the electron
temperature and mass) plasma with circularly polarized normal modes. In
general, however, both the plasma modes and radiation have
elliptical polarizations giving rise to generalized Faraday rotation.

\begin{figure}
\plotone{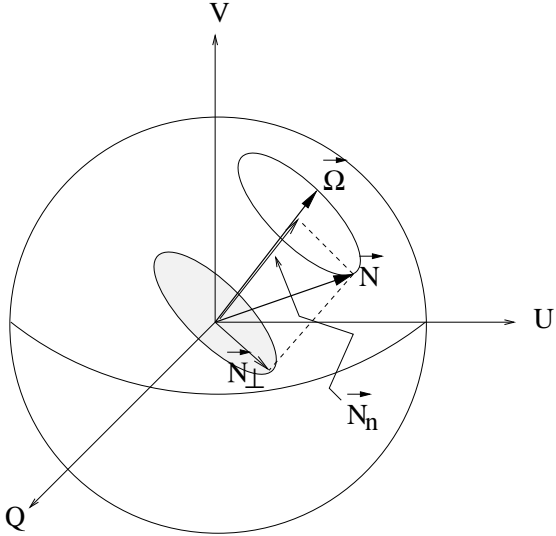}
\caption{The \poincare\ sphere is the unit sphere in a space defined
  by the three orthogonal Stokes parameters normalized by the total
  intensity $I$. Any polarized radiation can be
  described by a point on this sphere with the
  poles and equator denoting pure circular or linear polarization,
  respectively. The vector $\vec{\Omega}$ shows one of the normal modes of a magnetized plasma. Polarized light, denoted by
  $\vec{N}$, will rotate around $\vec{\Omega}$ at a constant angle
  as it passes through the plasma. Classic Faraday rotation
  corresponds to the case when the modes are coincident with the
  poles.}
\label{fig:sphere}
\end{figure}

An effective way to visualize the result of this general case is
to construct a three-dimensional space using the Stokes parameters as the Cartesian coordinates $(Q,U,V)/I$. The unit sphere in this space is the  
\poincare\ sphere \citep[e.g.,][]{km98,rb02} and is shown in Figure~\ref{fig:sphere}.
The surface of this sphere denotes all possible polarizations of the
radiation and plasma wave modes. The poles and the equator of the sphere
indicate either pure
circular or linear polarization, respectively. The two
orthogonal normal modes of a plasma are found as a line passing through
the center of the sphere and one mode is drawn as the vector $\vec{\Omega}$ in
Figure~\ref{fig:sphere}. A polarized light ray $\vec{N}$
passing through this medium will rotate around the wave mode at a constant
angle (shown in Fig.~\ref{fig:sphere} as the circle on the sphere's
surface) changing its polarization properties. The amount of
this generalized Faraday rotation depends on the relative orientation
between the polarization vector of the
propagating radiation and of the normal modes of the intervening
medium. The precession of the polarization vector can be analyzed by
decomposing it into one component, $\vec{N}_{\rm n}$, that is parallel
to $\vec{\Omega}$ and one component, $\vec{N}_{\perp}$ that is
perpendicular to it. The parallel component is unaffected by the
propagation through the medium, whereas the perpendicular component
undergoes rapid rotation on the plane perpendicular to
$\vec{\Omega}$, which results in a cyclic conversion between
linear and circular polarization \citep{pac73}. This mechanism for generating circular polarization has
long been considered as a possible origin for the observed circular
polarization from radio quasars \citep[e.g.,][]{rb02}. It also has been applied to \sgra\ in the context of a jet
emission model \citep{bf02}.

In a cold plasma the wave modes are coincident
with the poles, so polarized radiation will rotate around a line of
constant latitude on the sphere resulting in the classical Faraday
rotation. If such a cold plasma produces a large number of rotations
around the normal modes (i.e., it has a large rotation measure), then the radiation may
be linearly depolarized as the linear polarized vectors are averaged out over
different lines of sight. Interestingly, this does not occur for
generalized Faraday rotation. Even if the perpendicular component of the
polarization vector suffers very large rotations around
$\vec{\Omega}$, the emerging radiation will still have a net elliptical
polarization described by the component $\vec{N}_{\rm n}$ of the
incident radiation parallel to the normal mode. Therefore, there will
always remain some linearly polarized component to the radiation and it is not generally true that a
large rotation measure will lead to a linearly depolarized signal.

To begin to identify the polarization
signatures of a RIAF we first must calculate the ellipticity of the normal
modes in the accretion flow. The axial ratio of the elliptical wave modes in a plasma can be written as
\citep[e.g.,][]{mel97}:
\begin{equation}
\label{eq:T}
T_{\pm} = {(\alpha^{11} - \alpha^{22}) \mp \left [(\alpha^{11} -
    \alpha^{22})^2 + 4\alpha^{12} \alpha^{21} \right]^{1/2} \over
    {2i\alpha^{12}}},
\end{equation}
where $\epsilon^{ij}\equiv \delta^{ij} + (4\pi c /\omega^2)
\alpha^{ij}$ is the plasma dielectric tensor and $\alpha^{12} = -\alpha^{21}$. The components of $\alpha^{ij}$ are functions of $\omega \equiv
2\pi \nu$, where $\nu$ is the frequency of the propagating
radiation. The general expression for $\epsilon^{ij}$ for a thermal magnetized
plasma of any temperature was derived by \citet{tru58},
and can be expressed in an analytic form in either the very
low temperature ($\theta \ll 1$) or
very high temperature ($\theta \gg 1$) limit. If $T=0$ or $\infty$,
then the wave modes are linear and lie along the equator of the
\poincare\ sphere, and significant conversion between linear and circular polarization will occur. If $T=\pm1$ then the modes are circular and only standard
Faraday rotation is possible. 

Examination of equation~\ref{eq:T} shows that the key parameter in
determining the axial ratio of the linear modes is $(\alpha^{11} -
\alpha^{22})/\alpha^{12}$. For a plasma that has both a cold component with
number density $n_{\mathrm{cold}}$ and a relativistic ($\theta \gg
1$) component with density $n_{\mathrm{rel}}$ then \citep{mel97}
\begin{equation}
\label{eq:a11minusa22}
(\alpha^{11} - \alpha^{22}) = {-\omega_{\mathrm{B}}^2 c^2 \sin^2\phi
  e^2 \over m \omega^2} \left (n_{\mathrm{cold}} + 12\theta
  n_\mathrm{rel} \right)
\end{equation}
and
\begin{equation}
\label{eq:a12}
\alpha^{12}={i \cos\phi c^2 \omega_{\mathrm{B}} e^2 \over m \omega}
\left ( n_{\mathrm{cold}} + {n_{\mathrm{rel}} \over 2 \theta^2}\ln
\theta \right),
\end{equation}
where $\phi$ is the angle between the magnetic field and the direction
of propagation, and $\omega_{\mathrm{B}}\equiv eB/mc$ is the cyclotron frequency of an
electron in a magnetic field of strength $B$. These two expressions
are valid only for $\omega \gg \omega_{\mathrm{B}}$. For a
non-relativistic plasma, the wave modes are independent of temperature,
so the $\theta$ in equations~\ref{eq:a11minusa22} and~\ref{eq:a12} refers
to the temperature of the relativistic plasma. 

Consider now a thermal plasma with a temperature $kT_e$ that can
range from non-relativistic to ultra-relativistic, such as within an
RIAF \citep[e.g.,][]{yqn03}. We can use the
above equations to determine how the wave modes of this plasma will
vary with temperature and thus its effects on polarized radiation. Without loss of
generality, the distribution of energies in this
plasma can be written using the relativistic Maxwellian:
\begin{equation}
\label{eq:maxwell}
f(\gamma)d\gamma={\gamma^2 \beta e^{-mc^2/kT_e} d\gamma \over (kT_e/mc^2)
  K_2(mc^2/kT_e)}, 
\end{equation}
where $\gamma$ is the Lorentz factor, $\beta$ is the electron velocity
in units of $c$, and $K_2$ is the modified Bessel function. For any
value of the electron temperature, a critical Lorentz factor
$\gamma_{\mathrm{crit}}$ can be defined, above which
a fraction of this plasma will always be considered
ultra-relativistic. Integrating equation~\ref{eq:maxwell} above
$\gamma_{\mathrm{crit}}$ allows us to split the thermal plasma into a
relativistic component with density $n_{\mathrm{rel}}$ and a
cold component with density
$n_{\mathrm{cold}}=1-n_{\mathrm{rel}}$. Re-defining $\theta=\left < E
\right >/mc^2$, where $\left < E \right >$ is the mean energy of the
plasma above $\gamma_{\mathrm{crit}}$, the three quantities $(\alpha^{11} -
\alpha^{22})$, $\alpha^{12}$, and $T$ can all be calculated as
functions of the electron temperature.

\begin{figure}
\plotone{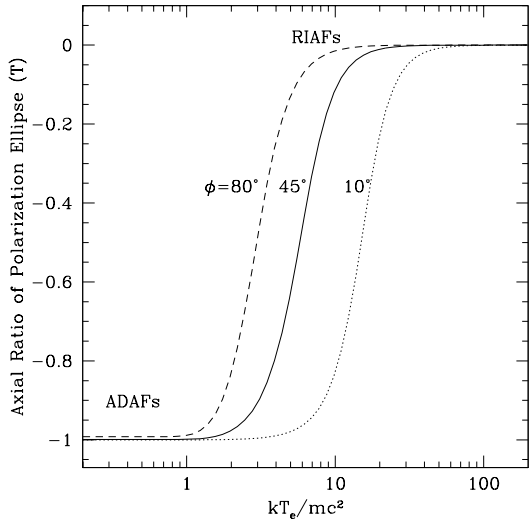}
\caption{Axial ratio of the polarization ellipse ($T$) defined by the normal
modes of a magnetized thermal plasma with temperature $kT_e$. A ratio
of $-1$ indicates modes which are purely circular (i.e., they lie
along the poles of the \poincare\ sphere), while a value of $0$ denotes
modes which are entirely linear (i.e., they lie along the equator of
the \poincare\ sphere). The plot shows the value of $T$ assuming an
observed frequency of 100~GHz and a magnetic field of 100~G. A
critical Lorentz factor of $\gamma_{\mathrm{crit}}=10$ was used to
define the ultra-relativistic electron population. The three curves
differentiate three assumptions for the angle between the
magnetic field and the direction of propagation, $\phi$.}
\label{fig:vskt}
\end{figure}

Figure~\ref{fig:vskt} shows the result of such a calculation and plots
the axial ratio of the normal modes
$T$ as a function of plasma temperature for an observed frequency of
100~GHz and a magnetic field of 100~G.
A value of
$\gamma_{\mathrm{crit}}=10$ was used for this and all subsequent calculations.
While approximations have been made to obtain this result, we see the
expected behavior that the modes are purely circular ($T=-1$) for
temperatures $\lesssim mc^2$. It is in this regime that ADAFs reside
\citep[e.g.,][]{opn00} and so they will only cause classical Faraday
rotation. The solid line in Figure~\ref{fig:vskt} shows the value of $T$
assuming a magnetic field that is tilted at $\phi=45^{\circ}$ to the
direction of propagation, while the dashed line shows the results when
$\phi=80^{\circ}$ and the dotted line denotes
$\phi=10^{\circ}$. At low values of the electron temperature, the magnetic field
geometry is practically irrelevant with the modes remaining nearly
purely circular.

In contrast, for a thermal plasma with a temperature $kT_e \gtrsim
mc^2$, the modes are elliptical resulting in generalized Faraday rotation. When the
temperature climbs to values so that $kT_e/mc^2 \sim
\gamma_{\mathrm{crit}}$, the wave modes become linear ($T=0$)
and radiation may undergo significant conversion between linear and
circular polarization. The magnetic field geometry has a large influence
when the modes are elliptical. When the field is closely aligned to
the direction of propagation, $\phi$ is small and the modes remain circular up
to larger temperatures, increasing the importance of Faraday
rotation. However, when the field is nearly perpendicular to the
direction of propagation, $\phi$ is large, increasing the ellipticity of the
wave modes and the importance of conversion. Both semi-analytic and
numerical simulations of RIAFs models predict relativistic electron
temperatures in the inner part of the accretion flows \citep{yqn03,sha07}. Therefore, the
wave modes will be elliptically polarized and generalized Faraday rotation will be important.

\section{Application to Models of \sgra}
\label{sect:sgra}
Linear polarization has been observed from \sgra\ in the
sub-millimeter range of the spectrum \citep{ait00,bow05,mar06}. At these frequencies, the
emission originates from synchrotron emitting thermal electrons
\citep[e.g.,][]{opn00,yqn03}. Since the emission is optically thick
and the synchrotron emissivity is a very steep function of radius \citep{lw07}, each
frequency can be mapped to a specific emission radius in the accretion
flow. Sub-millimeter VLBI observations have resolved the image of \sgra\ at a few
wavelengths \citep{k06}, allowing a tentative calibration of the
radius-frequency relation \citep{lw07}: $\nu_{11} = 1.0r_{13}^{-1}$, where
$\nu_{11}=\nu/(10^{11}\ \mathrm{Hz})$ and $r_{13}=r/(10^{13}\
\mathrm{cm})$, and a distance of 8~kpc has been assumed to \sgra.
\citet{lw07} have used this relation and the observed spectrum of
\sgra\ to estimate the run of the electron temperature and magnetic
field strength with radius: $kT_e/mc^2
\approx 14.9(2f_g)^{-1}r_{13}^{-0.4}$ and $B \approx 27 (2f_g)^2
r_{13}^{-0.2}$~G, where $f_g$ is a factor of order unity which we take
to be $f_g=1/\sqrt{2}$. These values are comparable with those found
from RIAF model fits to the \sgra\ spectrum \citep{yqn03}. 

\begin{figure}
\plotone{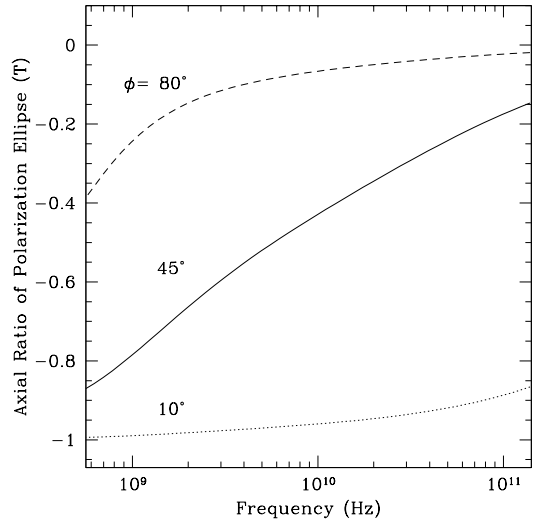}
\caption{Axial ratio of the polarization ellipse ($T$; eq.~\ref{eq:T})
  at the photosphere for each frequency $\nu$ in a model of the
  RIAF onto \sgra. The three lines differentiate the results among
  three different magnetic field geometries.}
\label{fig:vsnu}
\end{figure}

With these three relations, we can use the procedures described in
\S~\ref{sect:plasma} to calculate the axial ratio of the normal modes
$T$ as a function of the radial distance from the black hole. This is equivalent to
calculating $T$ at the emission point for each frequency
$\nu$. We plot in Figure~\ref{fig:vsnu} the results of this exercise, where
we have calculated $T$ from 6 to $\sim 1500$~Schwarzschild radii, which translates to
frequencies from $\sim 1$--$100$~GHz, for three different magnetic
field geometries.
We have assumed a black hole
mass of $4\times 10^6$~M$_{\odot}$ \citep{ghez03}, spherical geometry, and have ensured
that $\omega \gg \omega_{\mathrm{B}}$ at every radius.
The gas temperature in this model of \sgra\ is large enough over the
entire range of radii that the plasma normal modes are nearly always
highly elliptical and thus generalized Faraday rotation will be relevent. This plot cannot give a specific
prediction for the amount of circular and linear polarization expected
from \sgra\ --- that must wait for a full polarized radiative transfer
calculation through an MHD accretion simulation --- but it does
indicate that conversion will be important in the millimeter and
sub-millimeter, and, thus if RIAF models are accurate, circular
polarization is likely to be observed at these wavelengths.

As we saw in the previous section, the magnetic field geometry in the
accretion flow will also have an effect on the shape of the linear
modes. Figure~\ref{fig:vsnu} shows that a line-of-sight nearly aligned
with the magnetic field will significantly suppress conversion to
circular polarization over the entire frequency range. Of course,
polarization observations include the entire accretion flow within the
beam, therefore, for a purely radial magnetic field, only the central
region of the beam will observe lines-of-sight with low values of
$T$. However, numerical simulations of accretion flows find that the
velocity shear in the gas stretches radial magnetic fields
in the azimuthal direction \citep[e.g.,][]{hk01}, generating a
toroidal geometry (i.e., a large $\phi$). Figure~\ref{fig:vsnu} shows
that, in this case, conversion between linear and circular polarization will be
enhanced in the beam center. It seems possible, therefore, that
measurements of the percentage of both linear and circular
polarization over a range of frequency may allow a constraint to be
placed on the overall magnetic field geometry in the accretion flow.

These results clearly indicate that generalized Faraday rotation will
be important in RIAFs, and thus could impact the interpretation of
the linear polarization observations of \sgra. As discussed in
\S~\ref{sect:plasma}, if the generalized Faraday rotation is large
enough, then the observed linear polarization is just the projected
polarization of the plasma normal modes. Changes in the measured
position angle with frequency are then due to differences in the plasma
normal modes at different radii, and not to classical Faraday
rotation. Thus, unless the generalized Faraday rotation is very small, the
constraints on the accretion rate onto \sgra\ derived
from the classical rotation measure \citep{mar07} will not be
valid. If this is the case and the linear polarization measurements
track the changes in the plasma normal modes, then these observations
remain powerful probes of the accretion flow.

The importance of generalized Faraday rotation in the accretion flow
onto \sgra\ can be estimated by calculating the angle through which
the polarization vector will rotate \citep{mel97}
\begin{equation}
\Delta \psi = {12 \pi e^4 \over \omega^3 m^3 c^3} \int
n_{\mathrm{rel}} \theta B^2 \sin^2 \phi ds,
\label{eq:dpsi}
\end{equation}
where we have assumed a purely relativistic gas. Taking representative
values and assuming $ds \approx r$ we obtain
\begin{eqnarray}
\Delta \psi & \sim & 16 \left ( {r \over 10^{13}\ \mathrm{cm}} \right )
\left( n_{\mathrm{rel}} \over 10^6\ \mathrm{cm}^{-3} \right ) \left ( B \over 20\
\mathrm{G} \right )^2 \left ( \theta \over 10 \right) \times \nonumber \\ 
 & & \left ( \omega
\over 2\pi(10^{11}\ \mathrm{Hz}) \right )^{-3} \sin^2 \phi\ \mathrm{rads}.
\label{eq:dpsi2}
\end{eqnarray}
Therefore, it seems plausible that generalized Faraday rotation will
be important in \sgra, and the observed linear polarization
measurements must be carefully interpreted. More precise calculations
of $\Delta \psi$ will be performed in future work.

\section{Conclusions}
\label{sect:concl}
The latest RIAF models of \sgra\ indicate that the thermal plasma is
relativistic over a large range of radii \citep{yqn03,lw07}. This
paper has shown that, in general, RIAFs have highly elliptical wave
modes and therefore generalized Faraday rotation will dominate at millimeter and sub-millimeter
wavelengths. Since generalized
Faraday rotation in RIAFs will not depolarize the linear polarized
component of the radiation even if the number of rotations is very large,
linear polarization observations of \sgra\ and other low-luminosity
accretion flows may directly probe the plasma normal modes, an
important diagnostic of the accretion physics.  In
addition, the elliptical normal modes of the plasma will cause a
cyclic conversion between linear and circular polarization in this
frequency range. As the ellipticity of the normal modes depend on the
magnetic field geometry, circular and linear polarization observations
of \sgra\ over a wide range of frequency may differentiate between
radial and toroidal magnetic field geometry in the accretion flow. Exact
predictions of these effects require a detailed numerical simulation
which is planned for future work. It is hoped that the general
considerations presented here will help motivate future linear and
circular polarization observational campaigns on \sgra.

\acknowledgments

DRB is supported by the University of Arizona Theoretical Astrophysics
Program Prize Postdoctoral Fellowship. 

{}

\end{document}